\documentclass[12pt]{article}
\usepackage{amsmath,amssymb}
\newcommand{\be}{\begin{equation}}
\newcommand{\ee}{\end{equation}}
\newcommand{\bea}{\begin{eqnarray}}
\newcommand{\eea}{\end{eqnarray}}

\def \th {$\theta_{\mu\nu}$}
\def \o {$\hat o_\mu$}
\def\P{Poincar\'e }

\begin{document}
\renewcommand {\theequation}{\thesection.\arabic{equation}}
\renewcommand {\thefootnote}{\fnsymbol{footnote}}
\vskip1cm
\begin{flushright}
\end{flushright}
\vskip1cm
\begin{center}
{\large\bf An Interpretation of Noncommutative Field Theory\\in
Terms of a Quantum Shift} \vskip1cm {\bf M. Chaichian$^1$, K. Nishijima$^{1,2}$ and A. Tureanu$^1$}\\
$^1$High Energy Physics Division, Department of Physical Sciences,\\
University of Helsinki, and Helsinki Institute of Physics\\ P.O. Box
64, FIN-00014
Helsinki, Finland\\
$^2$Department of Physics, University of Tokyo 7-3-1 Hongo,\\
Bunkyo-ku, Tokyo 113-0033, Japan
\end{center}
\vskip1cm
\begin{abstract}

Noncommutative coordinates are decomposed into a sum of geometrical
ones and a universal quantum shift operator. With the help of this
operator, the mapping of a commutative field theory into a
noncommutative field theory (NCFT) is introduced. A general measure
for the Lorentz-invariance violation in NCFT is also derived.
\end{abstract}
\vskip1cm

\section{Introduction}
NCFT is characterized by the NC space-time coordinates $\hat x_\mu$
satisfying the commutation relations (CR)
\be\label{cr}[\hat x_\mu,\hat x_\nu]=i\theta_{\mu\nu}\ ,\ee
where \th\ is an antisymmetric constant matrix (for reviews, see,
e.g., \cite{SW,DN,Szabo}). In accordance with the contemporary
wisdom we may assume that the time-space noncommutativity is absent
(due to the violation of unitarity and causality \cite{GM,SST}),
\be\theta_{0j}=0\ ,\ \ \ \ j=1,2,3\ .\ee
Then, in order to offer a physical interpretation of (\ref{cr}), we
shall decompose $\hat x_\mu$ as follows:
\be\label{decomp}\hat x_\mu=x_\mu+\hat o_\mu\ ,\ee
where $x_\mu$ denotes the classical geometrical coordinates and
$\hat o_\mu$ a quantum-mechanical fluctuation, to be referred to as
a quantum shift operator hereafter.

Then the CR (\ref{cr}) can be reproduced by postulating a more
universal one,
\be\label{cr'}[\hat o_\mu,\hat o_\nu]=i\theta_{\mu\nu}\ .\ee
Because of the uniformity of space-time we may assume that this
quantum shift is common to all the classical positions, so that we
may also decompose $\hat y_\mu$, different from $\hat x_\mu$, as:
\be\hat y_\mu=y_\mu+\hat o_\mu\ .\ee
Then let us consider a function of $x_\mu$, say $f(x)$, and quantize
$x$ as $f(x)\rightarrow f(\hat x)$, and we may express this process
as follows by making use of the Taylor expansion:
\be\label{taylor} f(x)\rightarrow f(\hat x)=f(x+\hat o)=e^{\hat
o\cdot\partial}f(x)\ ,\ee
where
\be\hat o\cdot\partial=\hat o_\mu\partial^\mu\ .\ee
This operation is essentially a translation in space by $\hat o$,
the quantum shift.

We have to point out that for defining an expression such as
(\ref{taylor}), $f(x)$ has to be a linear combination of the
eigenfunctions of $\partial_\mu$. This happens to be the Fourier
representation of $f(x)$ and we can utilize the relation
\be e^{\hat o\cdot\partial}e^{ikx}=  e^{\hat o\cdot k}e^{ikx}=
e^{ik(x+\hat o)}=e^{ik\hat x}\ .\ee
When $f(x)g(x)=h(x)$ we generally get
$$
f(\hat x)g(\hat x)\neq g(\hat x)f(\hat x)
$$
and
$$
f(\hat x)g(\hat x)\neq h(\hat x)
$$
as we will show in Section 3.

\section{The Weyl-Moyal product}

In quantum mechanics we introduce a Hilbert space ${\cal
H}_{{phys}}$, but the quantum mechanical operator \o\ implies the
introduction of a {\it new} independent Hilbert space ${\cal H}_s$,
so that we are led to an extended Hilbert space ${\cal H}$, defined
by the direct product of these two spaces:
\be{\cal H}={\cal H}_{{phys}}\otimes{\cal H}_s\ . \ee
The quantum shift is realized, as it has been shown in
(\ref{taylor}) by applying the unitary operator
\be\label{unit} e^{\hat o\cdot\partial}\ee
to the wave function. This operation is a sort of phase
transformation representing a parallel translation in space, so that
it cannot be recognized for a single-particle system.

Now we shall proceed to a two-particle system represented by the
wave function $f(x_1)g(x_2)$. Then its quantum shift is realized as
\bea f(x_1)g(x_2)&\rightarrow& f(\hat x_1)g(\hat x_2)\cr &=& e^{\hat
o\cdot\partial^1}e^{\hat o\cdot\partial^2}f(x_1)g(x_2)\eea
in an obvious notation. The application of the
Baker-Campbell-Haussdorff formula in this case yields
\be e^{\hat o\cdot\partial^1}e^{\hat o\cdot\partial^2}=e^{\hat
o\cdot(\partial^1+\partial^2)}e^{\frac{i}{2}(\theta\partial^1\partial^2)}\
,\ee
where
\be \theta\partial^1\partial^2\equiv
\theta^{\mu\nu}\partial^1_\mu\partial^2_\nu\ .\ee
Thus we arrive at the following relationship
\be\label{qprod} f(\hat x_1)g(\hat x_2)= e^{\hat
o\cdot(\partial^1+\partial^2)}f(x_1)\star g(x_2)\ ,\ee
where
\be\label{sp}f(x_1)\star
g(x_2)=e^{\frac{i}{2}(\theta\partial^1\partial^2)}f(x_1)g(x_2)\ .\ee
This generalizes the Weyl-Moyal  $\star$-product, which is obtained
for $x_1=x_2=x$:
\be f(x)\star
g(x)=e^{\frac{i}{2}(\theta\partial^1\partial^2)}f(x_1)g(x_2)|_{x_1=x_2=x}\
.\ee
In (\ref{qprod}) the front factor corresponds to an unobservable
phase transformation depicting a parallel translation of the centre
of mass by $\hat o$ and the Weyl-Moyal product reflects the
fluctuation of the relative coordinates of the two particles of this
system. Eq. (\ref{sp}) can be generalized to a product of $n$
functions:
\bea\label{general} f_1(\hat x_1)...f_n(\hat x_n)&=&e^{\hat
o\cdot(\partial^1+...+\partial^n)}f_1(x_1)\star...\star f_n(x_n)\cr
&=&e^{\hat o\cdot(\partial^1+...+\partial^n)}e^Df_1(x_1)...
f_n(x_n)\ ,\eea
where
\be\label{D}
D=\frac{i}{2}\sum_{a<b}\theta^{\mu\nu}\partial^a_\mu\partial^b_\nu\
. \ee

Next, we introduce the space integral of a product of functions of
$\hat x$ and assume that the integral is {\it translationally
invariant} in $x$. Then we have
\bea\label{int}\int d^3xf_1(\hat x)...f_n(\hat x)&=&\int d^3x
e^{\hat o\cdot\partial^x} f_1(x)\star...\star f_n(x)\cr &=&\int d^3x
f_1(x)\star...\star f_n(x)+\int d^3x\hat
o^\mu\partial_\mu\left(f_1(x)\star...\star
f_n(x)\right)+...\cr&=&\int d^3x f_1(x)\star...\star f_n(x)\ . \eea
We realize that the final result is independent of the quantum shift
operator $\hat o$.

The space-integral of Moyal products has various properties relevant
to the formulation of NCFT.

1) The integral (\ref{int}) is invariant under cyclic permutations
of the $n$ complex-valued functions
\be\label{permute}\int d^3x f_1(x)\star...\star f_n(x)=\int d^3x
f_2(x)\star...\star f_n(x)\star f_1(x)\ .\ee

2) In the Moyal product we can replace under the integral one of the
$\star$-products by a usual (dot) product:
\bea\label{dotprod}\int d^3x f_1(x)\star...\star f_n(x)&=&\int d^3x
f_1(x)\cdot(f_2(x)\star...\star f_n(x))\cr&=&\int d^3x (f_1(x)\star
f_2(x))\cdot(f_3(x)\star...\star f_n(x))\cr&=&... \ .\eea
We skip the proofs since they follow directly from (\ref{general}).

\section{Noncommutative Field Theory}

Let us assume that $f(x)\cdot g(x)=h(x)$. Then we readily recognize
the following mismatch:
\bea f(\hat x) g(\hat x)&=&e^{\hat o\cdot\partial} f(x)\star g(x)\cr
&\neq& e^{\hat o\cdot\partial} f(x)\cdot g(x)=h(\hat x)\ .\eea

Thus, it is important to specify uniquely the factorization of a
given function before introducing the quantum shift.

{\it Rule 1

In NCFT we decompose a given operator into a product of primitive
factors or a linear combination of such products and make the
replacement $x\rightarrow\hat x$ in the primitive factors, thereby
identifying incoming fields with primitive factors.}

In what follows we shall confine ourselves to a neutral scalar
theory, and $\phi (x)$ and $\Phi(x)$ shall denote the incoming field
and the Heisenberg field, respectively. Then we may write
\be\label{infield}\phi(\hat x)=\phi(x+\hat o)=e^{\hat
o\cdot\partial}\phi(x)\ .\ee
For the Heisenberg field, however, eq. (\ref{infield}) is not valid
and it should be modified as follows:
\be\label{heisenberg_fields} \Phi(\hat x)= \Phi(x+\hat o)=e^{\hat
o\cdot\partial}\Phi_\theta(x)\ ,\ee
in accordance with the {\it Rule 1}. As we shall see later,
$\Phi_\theta(x)$ denotes the NC field corresponding to $\Phi$, and
it satisfies a field equation obtained by modifying the one for the
commutative field $\Phi$.

%
%
%
%

\section{Action principle}

The Lagrangian density for the neutral scalar field with $\Phi^4$
interaction, in the conventional commutative field theory, is given
by
\be\label{lagr}{\cal
L}=-\frac{1}{2}\left(\partial^\mu\Phi\partial_\mu\Phi+m^2\Phi^2\right)-\frac{\lambda}{4!}\Phi^4\
.\ee
The action principle for commutative field theory is given by
\be\label{action_pr}\delta\int d^4x{\cal L}(x)=0\ .\ee
The corresponding action principle for the NCFT is given by
\be\label{nc_action_pr}\delta\int d^4x{\cal L}(\hat x)=0\ ,\ee
with the same postulated functional form of the Lagrangian density
as in (\ref{action_pr}).

In deriving the field equation we assume that the field is not yet
quantized and is treated as a complex function. Inside the
Lagrangian density we make the replacement
(\ref{heisenberg_fields}). Then, with the help of eqs.
(\ref{permute}) and (\ref{dotprod}), we find
\bea \int d^4x{\cal L}(\hat x)&=& -\frac{1}{2}\int d^4 x\
\left(\partial^\mu\Phi_\theta\partial_\mu\Phi_\theta+m^2\Phi_\theta^2\right)\cr
&-&\frac{\lambda}{4!}\int d^4 x\
\Phi_\theta(x)\star\Phi_\theta(x)\star\Phi_\theta(x)\star\Phi_\theta(x)\
.\eea
In taking the variation, the second integrand has a rather
unfamiliar form, so that we handle it separately.
\bea \delta\int
d^4x\Phi_\theta(x)\star\Phi_\theta(x)\star\Phi_\theta(x)\star\Phi_\theta(x)&=&
\int d^4 x \
\{\delta\Phi_\theta(x)\star\Phi_\theta(x)\star\Phi_\theta(x)\star\Phi_\theta(x)\cr
&+&\Phi_\theta(x)\star\delta\Phi_\theta(x)\star\Phi_\theta(x)\star\Phi_\theta(x)
+...\}\cr &=&4 \int d^4
x\delta\Phi_\theta(x)\cdot\left(\Phi_\theta(x)\star\Phi_\theta(x)\star\Phi_\theta(x)\right)\
,\cr\eea
where use has been made of (\ref{permute}) and (\ref{dotprod}). Thus
the field equation for $\Phi_\theta$ turns out to be given by
\be\label{eom}
(\Box-m^2)\Phi_\theta(x)-\frac{\lambda}{3!}\Phi_\theta(x)\star\Phi_\theta(x)\star\Phi_\theta(x)=0\
.\ee
It is clear the $\Phi_\theta$ satisfies the NC version of the
neutral scalar theory.

\section{The S-matrix}

The S-matrix is a functional of the incoming field $\phi(x)$,
relating the $|\mbox{in}\rangle$ and $|\mbox{out}\rangle$ states
\be |\mbox{out}\rangle=S |\mbox{in}\rangle \ee
and we have the NC version defined by
\be\label{s-amtrix}S[\phi(\hat x)]=S_\theta[\phi(x)]\ ,\ee
where $S_\theta$ denotes the S-matrix in the NCFT.

Dyson's formula for the S-matrix in commutative field theory is:
\be\label{dyson-exp}S[\phi(x)]=T
\exp\left[i\int_{-\infty}^{\infty}dt\int d^3x{\cal
L}^{\mbox{int}}(\vec x,t)\right]\ ,\ee
where ${\cal L}^{\mbox{int}}$ denotes the interaction Lagrangian
density in commutative field theory and is given by (\ref{lagr}) as
\be\label{int_lagr}{\cal
L}^{\mbox{int}}=-\frac{\lambda}{4!}\phi^4(x)\ .\ee
Now we make the replacement $x\rightarrow \hat x$ in
(\ref{int_lagr}) and integrate over the space coordinates
\bea\int d^3x{\cal L}^{\mbox{int}}(\hat x)&=&-\frac{\lambda}{4!}\int
d^3 x\ \phi(x)\star\phi(x)\star\phi(x)\star\phi(x)\cr &=&\int
d^3x{\cal L}^{\mbox{int}}_\theta(x)\ .\eea
Hence
\bea\label{nc_s} S_\theta[\phi(x)]&=&S[\phi(\hat x)]\cr &=&T
\exp\left[i\int_{-\infty}^{\infty}dt\int d^3x{\cal
L}^{\mbox{int}}_\theta(\vec x,t)\right]\ .\eea
This S-matrix is obtained by replacing ${\cal L}^{\mbox{int}}$ by
${\cal L}^{\mbox{int}}_\theta$.

\section{The Heisenberg Field}

The Heisenberg field $\Phi(x)$ in commutative field theory can be
expanded in powers of the coupling constant as
\bea \Phi(x)=\phi(x)&+&i\int_{-\infty}^t dt'\ \left[\phi(x),\int
d^3x'{\cal L}^{\mbox{int}}(\vec x',t')\right]\cr
&+&i^2\int_{-\infty}^t dt'\int_{-\infty}^{t'}dt''\
\left[\left[\phi(x), \int d^3x'{\cal L}^{\mbox{int}}(\vec
x',t')\right],\int d^3x''{\cal L}^{\mbox{int}}(\vec
x'',t'')\right]\cr &+&...\ . \eea
In NCFT we replace $x$ by $\hat x$ and obtain
\bea e^{\hat o\cdot\partial^x}\Phi_\theta(x)&=&e^{\hat
o\cdot\partial^x}\{\phi(x)+i\int_{-\infty}^{t}dt'\
\left[\phi(x),\int d^3x'{\cal L}^{\mbox{int}}_\theta(\vec
x',t')\right]\cr &+&i^2\int_{-\infty}^{t}dt'\int_{-\infty}^{t'}dt''\
\left[\left[\phi(x), \int d^3x'{\cal L}^{\mbox{int}}_\theta(\vec
x',t')\right],\int d^3x''{\cal L}^{\mbox{int}}_\theta(\vec
x'',t'')\right]\cr &+&...\}. \eea
Apparently $\Phi_\theta(x)$ denotes the Heisenberg field in the NCFT
characterized by the interaction Lagrangian density ${\cal
L}^{\mbox{int}}_\theta$.
%
%
%
%

\section{Violation of Lorentz Invariance}

NC QFT violates Lorentz invariance, however, it possesses twisted
Poincar\'e invariance \cite{CKT}. Consequently the individual
elementary fields are in the representations of the usual Poincar\'e
group, i.e., in the case of the scalar field:
\bea\label{lorentz_repr} [\phi(x),P_\mu]&=&{\cal P}_\mu\phi(x)\ ,\ \
\ \ \ \ \ {\cal P}_\mu=\frac{1}{i}\frac{\partial}{\partial x^\mu}\
,\cr [\phi(x),M_{\mu\nu}]&=&{\cal M}_{\mu\nu}\phi(x)\ ,\ \ \ \ {\cal
M}_{\mu\nu}=\frac{1}{i}\left(x_\mu\frac{\partial}{\partial
x^\nu}-x_\nu\frac{\partial}{\partial x^\mu}\right)\ . \eea
In commutative field theory, the invariance of the $S$-matrix under
the generators of the Lorentz group is expressed through the
commutator:
\be [S,M_{\mu\nu}]=0\ . \ee

In the noncommutative case,
\be\label{S-violation} [S_\theta,M_{\mu\nu}]\neq 0\  \ee
where $S_\theta$ is given by (\ref{nc_s}), i.e.
$$ S_\theta=T
\exp\left[i\int d^4x{\cal L}_\theta^{\mbox{int}}(x)\right]\ . $$
The nonvanishing expression which represents the commutator
(\ref{S-violation}) gives the amount of the violation of Lorentz
invariance in NC QFT and can be found as follows.

Explicitly, the commutator (\ref{S-violation}) can be written as:
\be\label{S-com} [S_\theta,M_{\mu\nu}]= T\left[i\int d^4 y[{\cal
L}_\theta^{\mbox{int}}(y),M_{\mu\nu}]\exp\left(i\int d^4x{\cal
L}_\theta^{\mbox{int}}(x)\right)\right]\ . \ee
One can chose for ${\cal L}_\theta^{\mbox{int}}(x)$ any $n$-linear
form :
\bea\label{Lagr} {\cal
L}_\theta^{\mbox{int}}(x)&=&\sum_{i_1...i_n}f_{i_1...i_n}\phi_{i_1}^1(x)\star...\star\phi_{i_n}^n(x)\cr
&=&e^D\sum_{i_1...i_n}f_{i_1...i_n}\phi_{i_1}^1(x_1)...\phi_{i_n}^n(x_n)|_{x_1=...=x_n=x}\cr&=&e^D{\cal
L}(x_1,...,x_n)|_{x_1=...=x_n=x}\ , \eea
with $i_j$, $j=1,...,n$ standing for spinorial or tensorial indices
and the coefficients $f_{i_1...i_n}$ chosen such as to make the
combination
\be{\cal
L}(x_1,...,x_n)=\sum_{i_1...i_n}f_{i_1...i_n}\phi_{i_1}^1(x_1)...\phi_{i_n}^n(x_n)\ee
{\it in the local limit} ($x_1=...=x_n=x$), a Lorentz scalar.

However, for the simplification of the argument, we shall take as an
illustration the NC $\lambda\phi^3$-theory, where
\be\label{nonloc}{\cal
L}(x_1,x_2,x_3)=\frac{\lambda}{3!}\phi(x_1)\phi(x_2)\phi(x_3)\ee
and, according to (\ref{D})
\be\label{D4}
D=\frac{i}{2}\theta^{\alpha\beta}\left[\partial^1_\alpha\partial^2_\beta
+\partial^1_\alpha\partial^3_\beta+\partial^2_\alpha\partial^3_\beta\right]\
.\ee
The action of the Lorentz generator on the composite object ${\cal
L}(x_1,x_2,x_3)$ makes use of the fact that the component fields
$\phi(x_i)$, $i=1,2,3$ are scalar representations of the Lorentz
group (see eq. (\ref{lorentz_repr})):
\bea [{\cal L}(x_1,x_2,x_3),M_{\mu\nu}]&=&\left({\cal
M}^{x_1}_{\mu\nu}\phi(x_1)\right)\phi(x_2)\phi(x_3)\cr&+&\phi(x_1)\left({\cal
M}^{x_2}_{\mu\nu}\phi(x_2)\right)\phi(x_3)\cr&+&\phi(x_1)\phi(x_2)\left({\cal
M}^{x_3}_{\mu\nu}\phi(x_3)\right)\cr&\equiv&{\cal
M}^{x_1x_2x_3}_{\mu\nu}{\cal L}(x_1,x_2,x_3)\ .\eea

Then we can compute the commutator of ${\cal
L}_\theta^{\mbox{int}}(y)$ with the Lorentz generator $M_{\mu\nu}$:
\bea [{\cal L}_\theta^{\mbox{int}}(y),M_{\mu\nu}]&=& e^{D}[{\cal
L}(y_1,y_2,y_3),M_{\mu\nu}]|_{y_1=y_2=y_3=y}\\&=&e^{D}{\cal
M}_{\mu\nu}^{y_1y_2y_3}{\cal L}(y_1,y_2,y_3)|_{y_1=y_2=y_3=y}\cr
&=&\left[e^{D}{\cal M}_{\mu\nu}^{y_1y_2y_3}e^{-D}\right]e^D{\cal
L}(y_1,y_2,y_3)|_{y_1=y_2=y_3=y}\cr &=&\left({\cal
M}_{\mu\nu}^{y_1y_2y_3}+[D,{\cal
M}_{\mu\nu}^{y_1y_2y_3}]\right)e^D{\cal
L}(y_1,y_2,y_3)|_{y_1=y_2=y_3=y}\cr &=&{\cal M}_{\mu\nu}^y{\cal
L}_\theta^{\mbox{int}}(y)+[D,{\cal M}_{\mu\nu}^{y_1y_2y_3}]e^D{\cal
L}(y_1,y_2,y_3)|_{y_1=y_2=y_3=y}\nonumber . \eea
In the expression of the commutator
\bea\label{D-com}[D,{\cal
M}_{\mu\nu}^{y_1y_2y_3}]&=&\frac{1}{2}\left[\sum_{1\leq
a<b\leq3}\theta^{\alpha\beta}\partial^a_\alpha
\partial^b_\beta,\sum_{k=1}^3(y^k_\mu\partial
^k_\nu-y^k_\nu\partial^k_\mu)\right]\\
&=&\frac{1}{2}\sum_{1\leq
a<b\leq3}\left(\theta^{\mu\beta}\partial^a_\nu\partial^b_\beta
-\theta^{\nu\beta}\partial^a_\mu\partial^b_\beta
+\theta^{\alpha\mu}\partial^a_\alpha\partial^b_\nu-\theta^{\alpha\nu}\partial^a_\alpha\partial^b_\mu\right)\nonumber\eea
one can view the action of ${\cal M}_{\mu\nu}^{y_1y_2y_3}$ on $D$ as
changing $\theta_{\nu\beta}\to\theta_{\mu\beta}$ (first term),
$\theta_{\mu\beta}\to-\theta_{\nu\beta}$ (second term),
$\theta_{\alpha\nu}\to\theta_{\alpha\mu}$ (third term) and
$\theta_{\alpha\mu}\to-\theta_{\alpha\nu}$ (last term). This amounts
to introducing and "auxiliary Lorentz generator", which transforms
properly $\theta_{\alpha\beta}$ as a tensor:
\be\label{D-com'}[D,{\cal M}_{\mu\nu}^{y_1y_2y_3}]=:{\cal
M}_{\mu\nu}^\theta D,\ee
with
\be[{\cal M}_{\mu\nu}^\theta,\theta_{\alpha\beta}]=
\frac{1}{i}(\eta_{\mu\alpha}\theta_{\nu\beta}-\eta_{\mu\beta}\theta_{\nu\alpha}-\eta_{\nu\alpha}\theta_{\mu\beta}+
\eta_{\nu\beta}\theta_{\mu\alpha})\ .\ee

In order to find a representation of ${\cal M}_{\mu\nu}^\theta$
(without the necessity of imposing the antisymmetry constraint on
the elements $\theta_{\alpha\beta}$), we define the matrix
$\sigma_{\alpha\beta}$ with totally independent components by
\be
\theta_{\alpha\beta}\equiv\sigma_{\alpha\beta}-\sigma_{\beta\alpha}\
, \ee
and re-express the above changes in terms of the $\sigma$-matrix:
\bea \sigma_{\nu\beta}&\to&\sigma_{\mu\beta}\ ,\ \ \ \
\sigma_{\mu\beta}\to-\sigma_{\nu\beta}\cr
\sigma_{\alpha\nu}&\to&\sigma_{\alpha\mu}\ ,\ \ \ \
\sigma_{\alpha\mu}\to-\sigma_{\alpha\nu}\ .\eea

Thus, we can represent ${\cal M}_{\mu\nu}^\theta$ as:
\be\label{new-op} {\cal
M}_{\mu\nu}^\theta=-\frac{1}{i}\left(\sigma_{\nu\beta}\frac{\partial}{\partial
\sigma_{\mu\beta}}-\sigma_{\mu\beta}\frac{\partial}{\partial
\sigma_{\nu\beta}}+\sigma_{\alpha\nu}\frac{\partial}{\partial
\sigma_{\alpha\mu}}-\sigma_{\alpha\mu}\frac{\partial}{\partial
\sigma_{\alpha\nu}}\right)\ . \ee

Hence (\ref{S-com}) yields
\be\label{S-com'}[S_\theta,M_{\mu\nu}]= T\left[i\int d^4
y\left({\cal M}_{\mu\nu}^y+{\cal M}_{\mu\nu}^\theta D\right){\cal
L}_\theta^{\mbox{int}}(y)\exp\left(i\int d^4x{\cal
L}_\theta^{\mbox{int}}(x)\right)\right]\ .\ee
Since
\bea\label{vanish} T\left[i\int d^4 y{\cal M}_{\mu\nu}^y{\cal
L}_\theta^{\mbox{int}}(y)\exp\left(i\int d^4x{\cal
L}_\theta^{\mbox{int}}(x)\right)\right]\cr =i\int d^4 y{\cal
M}_{\mu\nu}^y T \left[{\cal L}_\theta^{\mbox{int}}(y)\exp\left(i\int
d^4x{\cal L}_\theta^{\mbox{int}}(x)\right)\right]=0 \eea
and
\bea\label{abc} {\cal M}_{\mu\nu}^\theta{\cal
L}_\theta^{\mbox{int}}(y)&=&{\cal M}_{\mu\nu}^\theta e^D{\cal
L}(y_1,y_2,y_3)|_{y_1=y_2=y_3=y}\cr&=&({\cal M}_{\mu\nu}^\theta D)
e^D{\cal L}(y_1,y_2,y_3)|_{y_1=y_2=y_3=y}\cr&=&({\cal
M}_{\mu\nu}^\theta D){\cal L}_\theta^{\mbox{int}}(y)\ ,\eea
we conclude, by putting together (\ref{S-com'}) with (\ref{vanish})
and (\ref{abc}) and using (\ref{D-com'}), that
\be\label{LI-amount}[S_\theta,M_{\mu\nu}]={\cal M}_{\mu\nu}^\theta
S_\theta\ . \ee
The expression in the r.h.s. of (\ref{LI-amount}) represents in
general the amount of Lorentz-invariance violation of the $S$-matrix
in the case of NC QFT.

\section{Conclusions}

We may conclude that the NCFT is obtained from the corresponding
commutative field theory by a simple mapping resulting from the
replacement $x\rightarrow\hat x=x+\hat o$, which amounts to a
modification of ${\cal L}$ into ${\cal L}_\theta$. We should also
emphasize the fact that observable quantities depend only on
$\theta$ but never on $\hat o$ explicitly.

The violation of the Lorentz invariance in the action of NCFT has
been also found in general, in terms of a newly introduced operator,
or "auxiliary Lorentz generator", ${\cal M}_{\mu\nu}^\theta$ (eq.
(\ref{new-op})), whose role is to transform $\theta_{\alpha\beta}$
as a Lorentz tensor. Therefore, as expected, the theory would be
Lorentz invariant, had $\theta_{\alpha\beta}$ transformed properly
under Lorentz transformations. Though the expression of the
violation was derived in the particular case of the NC
$\lambda\phi^3$ theory, it is valid in general, since ${\cal
M}_{\mu\nu}^\theta$ acts solely on the $\theta$-variable, i.e. on
the $\star$-product, disregarding the actual fields involved in the
interaction, as long as the in-fields are in the representations of
the \P algebra, which are identical to the ones of the twisted
Poincar\'e.

\vskip 0.3cm {\bf{Acknowledgements}}. We would like to thank Peter
Pre\v{s}najder for useful discussions. One of the authors (K.N.) is
partially supported by Grant-in-Aid for Scientific Research from the
Ministry of Education, Culture, Sports, Science and Technology of
Japan. The financial support of the Academy of Finland under the
projects no. 54023 and 104368 is greatly acknowledged.

\end{document}